\documentclass[aps,prl,twocolumn,superscriptaddress,showpacs]{revtex4}

\usepackage{graphicx}

\begin{document}

\title{Effective Dissipation and Turbulence in Spectrally Truncated Euler Flows}

\author{Cyril Cichowlas}
\affiliation{Laboratoire de Physique Statistique de l'Ecole Normale 
Sup{\'e}rieure, \\
associ{\'e} au CNRS et aux Universit{\'e}s Paris VI et VII,
24 Rue Lhomond, 75231 Paris, France}
\author{Pauline Bona\" \i ti}
\affiliation{Laboratoire de Physique Statistique de l'Ecole Normale 
Sup{\'e}rieure, \\
associ{\'e} au CNRS et aux Universit{\'e}s Paris VI et VII,
24 Rue Lhomond, 75231 Paris, France}
\author{Fabrice Debbasch}
\affiliation{ERGA, CNRS UMR 8112, \\ 
4 Place Jussieu, \\ 
F-75231 Paris Cedex 05, France}
\author{Marc Brachet}
\affiliation{Laboratoire de Physique Statistique de l'Ecole Normale 
Sup{\'e}rieure, \\
associ{\'e} au CNRS et aux Universit{\'e}s Paris VI et VII,
24 Rue Lhomond, 75231 Paris, France}

\date{\today}

\begin{abstract}
ABSTRACT:
A new transient regime in the relaxation towards absolute equilibrium of the conservative and
time-reversible 3-D Euler equation
with high-wavenumber spectral truncation is characterized.
Large-scale dissipative effects, caused by the thermalized modes that spontaneously appear between a transition wavenumber and the maximum wavenumber, are calculated using fluctuation dissipation relations.
The large-scale dynamics is found to be similar to that of high-Reynolds number Navier-Stokes equations and thus to obey (at least approximately) Kolmogorov scaling.
\end{abstract}


\maketitle

Turbulence has been observed in inviscid and conservative
systems, in the context of (compressible) low-temperature superfluid turbulence \cite{Nor7,Nor4,Abid}.
This behavior has also been reproduced using simple (incompressible)
Biot-Savart vortex methods, which amount to Eulerian dynamics with {\sl ad hoc} vortex reconnection \cite{ATN02}. 
The purpose of the present letter is to study the dynamics of spectrally truncated 3-D incompressible Euler flows.
Our main result is that the inviscid and conservative Euler equation, with a high-wavenumber spectral truncation, has long-lasting transients which behave just as those of the dissipative (with generalized dissipation) Navier-Stokes equation. This is so because the thermalized modes between some transition wavenumber and the maximum wavenumber can act as a fictitious microworld providing an effective viscosity to the modes with wavenumbers below the transition wavenumber.

We thus study general solutions to the finite system of ordinary differential equations for the complex variables ${\bf \hat v}({\bf k})$ (${\bf k}$ is a 3 D vector of relative integers $(k_1,k_2,k_3)$ satisfying $\sup_\alpha |k_\alpha | \leq k_{\rm max}$)
\begin{equation}
{\partial_t { \hat v}_\alpha({\bf k},t)}  =  -\frac{i} {2} {\mathcal P}_{\alpha \beta \gamma}({\bf k}) \sum_{\bf p} {\hat v}_\beta({\bf p},t) {\hat v}_\gamma({\bf k-p},t)
\label{eq_discrt}
\end{equation}
where ${\mathcal P}_{\alpha \beta \gamma}=k_\beta P_{\alpha \gamma}+k_\gamma P_{\alpha \beta}$ with $P_{\alpha \beta}=\delta_{\alpha \beta}-k_\alpha k_\beta/k^2$ and the convolution in (\ref{eq_discrt}) is truncated to $\sup_\alpha |k_\alpha | \leq k_{\rm max}$, $\sup_\alpha |p_\alpha | \leq k_{\rm max}$ and $\sup_\alpha |k_\alpha-p_\alpha | \leq k_{\rm max}$.

This system is time-reversible and exactly conserves the kinetic energy 
$E=\sum_{k}E(k,t)$, where the energy spectrum $E(k,t)$ is
defined by averaging ${\bf \hat v}({\bf k'},t)$ on spherical shells of width $\Delta k = 1$,
\begin{equation}
E(k,t) = {\frac1 2} \sum_{k-\Delta k/2< |{\bf k'}| <  k + \Delta k/2} |{\bf \hat v}({\bf k'},t)|^2 \, .
\label{eq_energy}
\end{equation}

The discrete equations (\ref{eq_discrt}) are classically obtained \cite{Houches}
by performing a Galerkin truncation (${\bf \hat v}({\bf k})=0$ for 
$\sup_\alpha |k_\alpha | \leq k_{\rm max}$) on 
the Fourier transform  ${\bf v}({\bf x},t)=\sum {\bf \hat v}({\bf k},t) e^{i {\bf k}\cdot {\bf x}}$
of a spatially periodic velocity field obeying
the (unit density) three-dimensional incompressible Euler equations,
\begin{eqnarray}
{\partial_t {\bf v}}  + ({\bf v} \cdot \nabla) {\bf v}& =&- \nabla p   ~,  \nonumber \\
 \nabla  \cdot {\bf v} &=&0 ~.
 \label{eq_euler}
\end{eqnarray}
The short-time, spectrally-converged truncated Eulerian dynamics (\ref{eq_discrt}) has been studied \cite{FrischBlues,FDR-Cyril} to obtain numerical evidence for or against blowup of the original (untruncated) Euler equations (\ref{eq_euler}).
We will study here the behavior of solutions of (\ref{eq_discrt}) when spectral convergence to solutions of (\ref{eq_euler}) is lost.
Long-time truncated Eulerian dynamics is relevant to the limitations of standard simulations of high Reynolds number (small viscosity) turbulence which are performed using Galerkin truncations of the Navier-Stokes equation \cite{coursuriel}.

Equations (\ref{eq_discrt}) are solved numerically 
using standard \cite{Got-Ors} pseudo-spectral methods
with resolution $N$. 
The solutions are
dealiased by spectrally truncating the
modes for which at least one wave-vector component exceeds
 $N/3$
(thus a $1600^3$ run is truncated at $k_{\rm max}=534$).
This method allows the exact evaluation of the Galerkin convolution in (\ref{eq_discrt}) in only $N^3 \log{N}$ operations.
Time marching is done with a second-order
leapfrog finite-difference scheme, even and odd time-steps are periodically re-coupled
using fourth-order Runge-Kutta.

To study the dynamics of (\ref{eq_discrt}), we use the so-called Taylor-Green 
\cite{TG1937}  single--mode initial
condition  of (\ref{eq_euler})
${u}^{\rm TG}=\sin{x}\cos{y}\cos{z}$,
${v}^{\rm TG}=-{u}^{\rm TG}(y,-x,z)$,
${w}^{\rm TG}=0$.
Symmetries are employed in a standard way \cite{Brachet1} to reduce
memory storage and speed up computations.
Runs were made with $N=256$, $512$, $1024$ and $1600$.

\begin{figure}[ht!]
\begin{center}
 \hspace{-0.cm}  \includegraphics[width=9.5cm,height=11.cm,angle=0]{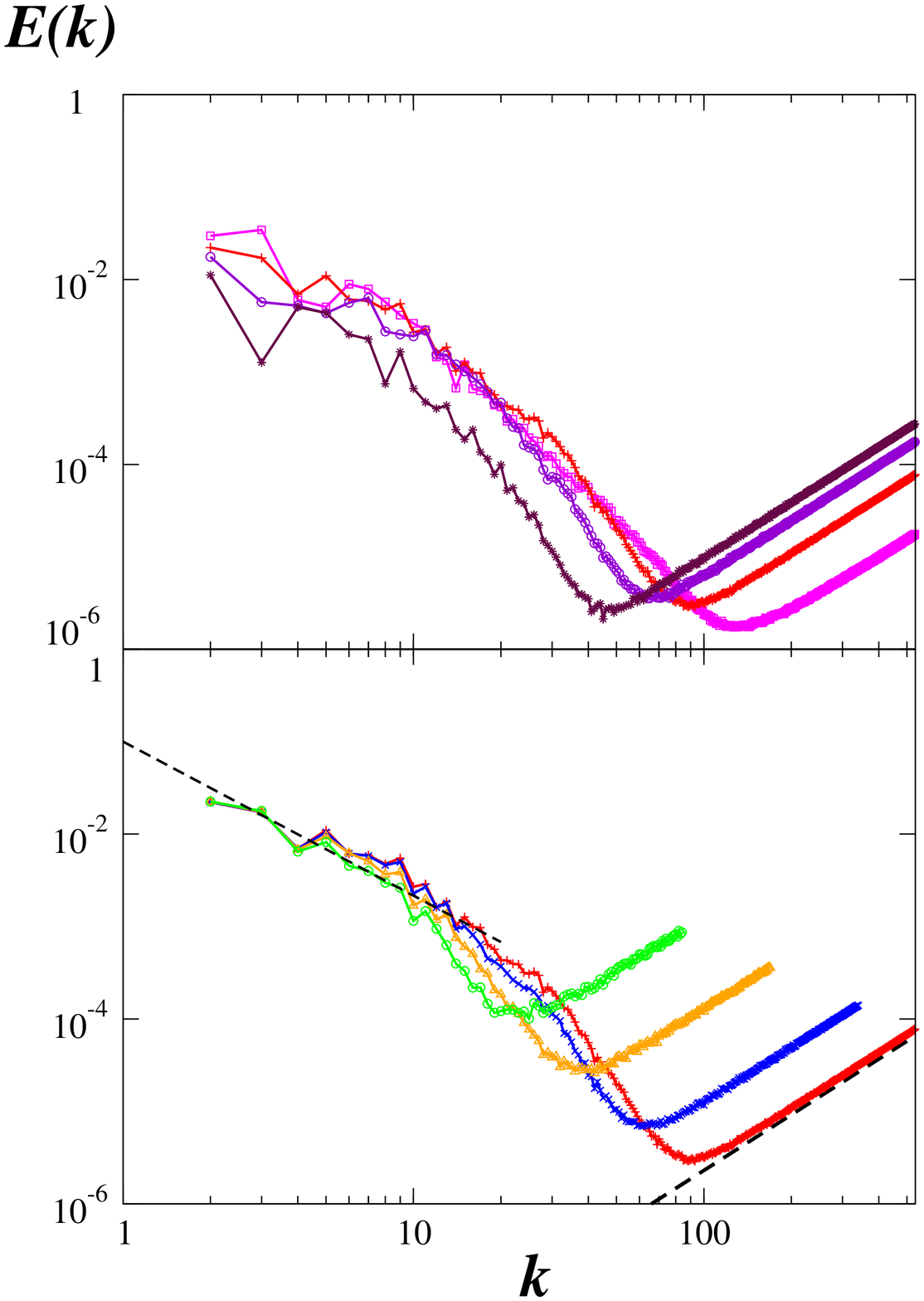} 
\vspace{-1.2cm} 
\caption{\small{Energy spectra, top: resolution $1600^3$ at $t = (6.5, 8,10,14)$ ($\diamond$, $ + $, $\circ$, $\ast$); bottom: resolutions $256^3$ (circle  $\circ$), $512^3$ (triangle $\triangle$), $1024^3$ (cross $\times$) and $1600^3$ (cross $+$) at $t=8$. The dashed lines indicate $k^{-5/3}$ and $k^2$ scalings.}
\label{SpecTh}}
\end{center}
\end{figure}

Figure \ref{SpecTh} displays the time evolution (top) and resolution dependence (bottom) of the energy spectra. 
Each energy spectrum $E(k,t)$ admits a minimum at $k=k_{\rm th}(t)<k_{\rm max}$, in sharp contrast with the short-time ($t\leq4$) spectrally converged Eulerian dynamics (data not shown, see \cite{Brachet1,FDR-Cyril}). For $k>k_{\rm th}(t)$ the energy spectrum obeys the scaling law $E(k,t)=c(t) k^2$ (see the dashed line at the bottom of the figure). 
The dynamics thus spontaneously generates a scale separation at wavenumber $k_{\rm th}(t)$. 
Figure \ref{SpecTh} also shows that $k_{\rm th}$ slowly decreases with time. 
For fixed $k$ inside the $k^2$ scaling zone $E(k,t)$ increases with time but $E(k,t)$ decreases with time for $k$ close (but inferior) to $k_{\rm th}(t)$.

The traditionally expected  \cite{KRA73,Houches} asymptotic dynamics of the system is to reach an absolute equilibrium, which is a statistically stationary exact solution of the truncated Euler equations, with energy spectrum $E(k)= c k^2$. 
Our new results (see figure \ref{SpecTh}) show that a time-dependent statistical equilibrium appears long before the system reaches its stationary state.
Indeed,  the early appearance of a $k^2$ zone is the key factor in the relaxation of the system towards the absolute equilibrium: as time increases, more and more modes gather into a time-dependent statistical equilibrium which itself tends towards an absolute equilibrium. 

\begin{figure}[ht!]
\begin{center}
 \hspace{-0.cm} \includegraphics[height=9.5cm,angle=-90]{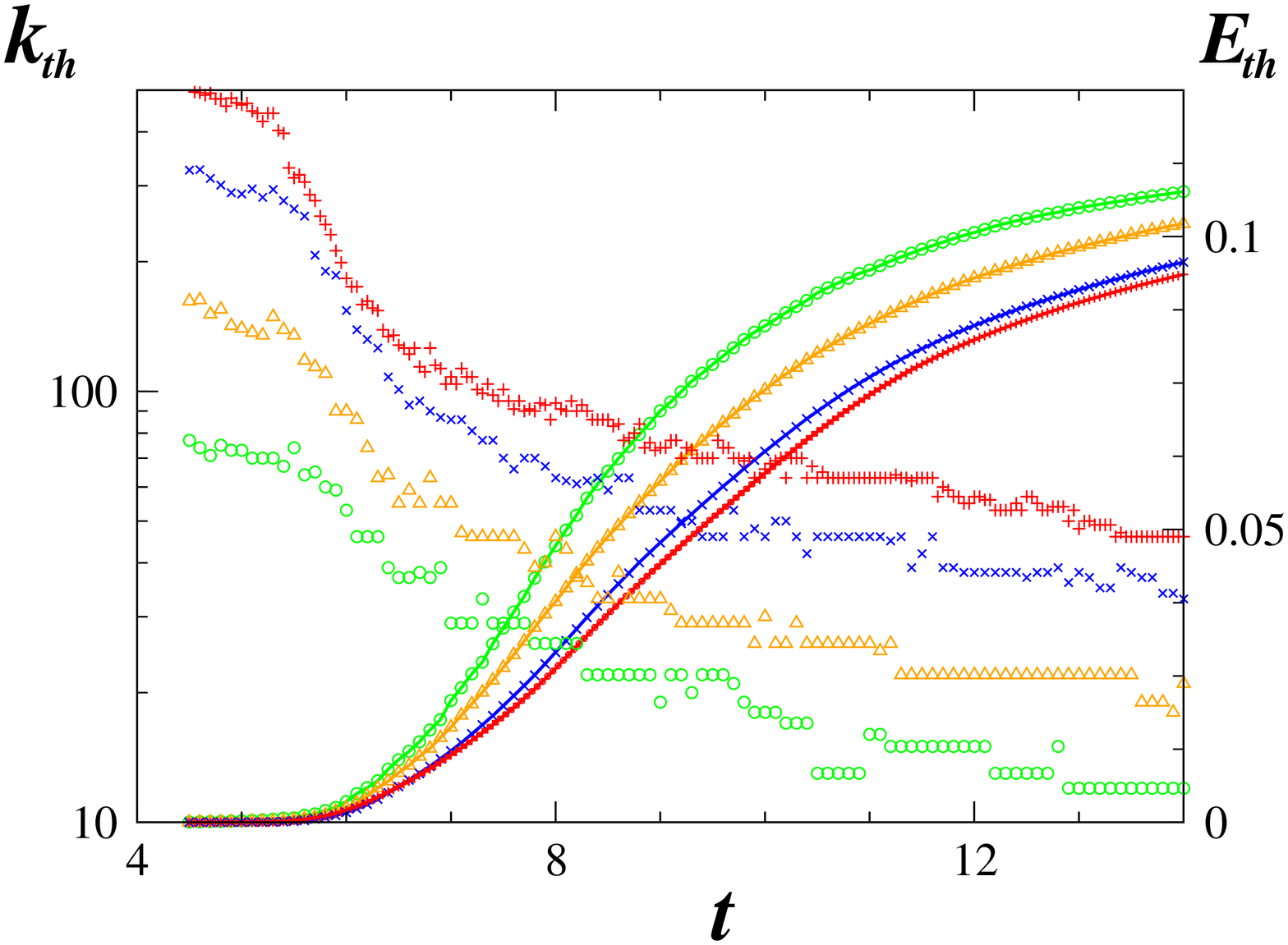} 
\vspace{-0.5cm} 
\caption{\small{Time evolution of $ k_{\rm th}$ (left vertical axis) and $E_{\rm th}$ (right vertical axis) at resolutions $256^3$ (circle  $\circ$), $512^3$ (triangle $\triangle$), $1024^3$ (cross $\times$) and $1600^3$ (cross $+$) .}
\label{tKmin}}
\end{center}
\end{figure}

Since the total energy $E$ is constant, the energy dissipated from large scales into the time dependent statistical equilibrium is given by
\begin{equation}
{E}_{\rm  th}(t) =  \sum_{k_{\rm th}(t) <  k } E(k,t) ~.
\label{Th_energy}
\end{equation}
The time evolutions of $k_{\rm th}$ and ${E}_{\rm  th}$ are presented on figure \ref{tKmin}. The figure clearly displays the long transient during which, for all resolutions, $k_{\rm th}$ decreases and ${E}_{\rm  th}$ increases with time. Note that, at all times, $k_{\rm th}$ increases and ${E}_{\rm  th}$ decreases with the resolution.

Since the energy of the 
time-dependent equilibrium increases with time, the modes outside the equilibrium lose
energy. The presence of a time-dependent equilibrium thus induces an effective dissipation on the lower $k$ modes.

We now estimate the characteristic time of effective dissipation $\tau_{\rm }(k_{\rm d})$
of modes $k_{\rm d}$ close to $k_{\rm th}(t)$
by assuming time-scale separation and studying, at each time $t$, the relaxation towards the {\it time-independent} absolute equilibrium characterized by ${E}_{\rm  th}(t)$ and $k_{\rm max}$.
The existence of a fluctuation dissipation
theorem (FDT) \cite{KRA59,CBDB-Big} ensures than dissipation around the equilibrium has the same characteristic time-scale as the equilibrium correlation functions 
$\left<
{\hat v}_\alpha ({\bf k},t) {\hat v}_\beta ({\bf k'},0) \right>
$
(brackets denote equilibrium statistical averaging
over initial conditions ${\hat v}_\beta ({\bf k'},0)$).
Defining this time scale $\tau_{\rm C}$ as the parabolic decorrelation time
\begin{equation} 
\tau_{\rm C}^2\partial_{tt} \left<
{\hat v}_\alpha ({\bf k},t) {\hat v}_\beta ({\bf k'},0) \right>_{| t=0}
  =
\left<
{\hat v}_\alpha ({\bf k},0) {\hat v}_\beta ({\bf k'},0) \right> ,
\label{taucdef}
\end{equation}
time translation invariance allows to express the second order time derivative as
$
-\left<
{\partial_{t}\hat v}_\alpha ({\bf k},t) \partial_{t'}{\hat v}_\beta ({\bf k'},t') \right>_{| t=t'=0}. 
$
Using expression (\ref{eq_discrt}) for the time derivatives
reduces the evaluation of $\tau_{\rm C}$
to that of an equal-time fourth-order moment of a gaussian field with correlation
$<{\hat v}_\alpha ({\bf k},t) {\hat v}_\beta ({\bf -k},t)>=A P_{\alpha \beta}({\bf k})$ \cite{Houches}
where
$A=E_{\rm th}/(2 k_{\rm max})^3$.
The only non-vanishing contribution is a one loop graph \cite{isserlis,coursuriel}.
The correlation time $\tau_{\rm C}$ associated to wavenumber $k$ is found in this way \cite{CBDB-Big} to obey the simple scaling law
\begin{equation}
\tau_{\rm C}=\frac{C}{k  \sqrt{E_{\rm th}}}  ~,
\label{MajdaCE}
\end{equation}
where $C=1.43382$ is a constant of order unity.
The time-scale $\tau_{\rm C}$ is the eddy turnover time at wavenumber $k_{\rm th}$. Because of Kolmogorov (K41) behavior (see below) the evolution of ${E}_{\rm  th}$ is governed by the large-eddy turnover time. The assumption of time-scale separation made above is thus consistent. 

\begin{figure}[ht!]
\begin{center}
 \hspace{-0.cm} \includegraphics[height=9.5cm,angle=-90]{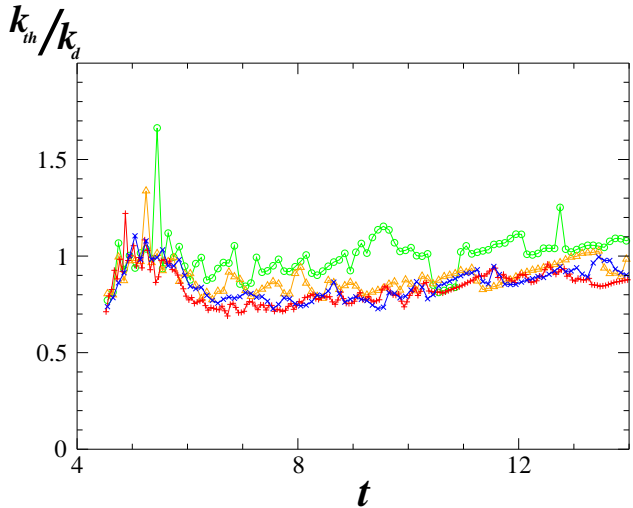}
\vspace{-0.2cm} 
\caption{\small{Time evolution of the ratio $k_{\rm th}/k_{\rm d}$  at resolutions $256^3$ (circle  $\circ$), $512^3$ (triangle $\triangle$), $1024^3$ (cross $\times$) and $1600^3$ (cross $+$) .}
\label{tkm}}
\end{center}
\end{figure}

This strongly suggests to introduce an effective generalized Navier-Stokes model for the dissipative dynamics of  modes $k$ close to $k_{\rm th}(t)$.
To wit, we make the  {\it Ansatz} $\varepsilon(k,t)=\bar{\nu}|k| E(k,t)$, where $\bar{\nu} =\sqrt{E_{\rm th}}/C$ and $\varepsilon(k,t) = -\partial E(k,t)/\partial t$ is the spectral density of energy dissipation
\begin{equation}
\varepsilon(t)=\frac{{\rm d} {E}_{\rm th}(t)}{{\rm d}t} ~.
\label{Eps_def}
\end{equation} 
Assuming that this dissipation takes place in a range of width $\alpha k_{\rm d}$ around $k_{\rm d}$ , we estimate
the total dissipation $\varepsilon \sim \bar{\nu} k_{\rm d} E(k_{\rm d}) \alpha k_{\rm d}$. This, together with $E(k_{\rm d})\sim k_{\rm d}^{2} {E}_{\rm th}/k_{\max}^3$ yields the relation
\begin{equation} 
k_{\rm d} \sim  \left( \frac{\varepsilon} {E_{\rm th}^{3/2}} \right)^{1/4} k_{\max}^{3/4} ~.
\label{eq:kd}
\end{equation}
The consistency of this estimation of effective dissipation with the results displayed in figure \ref{tKmin} requires that $k_{\rm d} \sim k_{\rm th}$.
The ratio $k_{\rm th}/k_{\rm d}$ is displayed on figure \ref{tkm}. It is seen to be of order unity and is reasonably constant in time and resolution independent (at least for $N>256$).

Thus the small-scale modes between $k_{\rm th}$ and $k_{\rm max}$ act as a fictitious thermostat providing, via the FDT, an effective viscosity to the large-scale modes with wavenumbers below $k_{\rm th}$.
Note that spontaneous equilibration
happening in conservative isolated systems, such as the one studied in the present letter, should not be confused with equilibration resulting from interaction with the thermalized degrees of freedom of the molecules constituting a physical fluid. Indeed the reversible dynamics of the isolated system (\ref{eq_discrt}) spontaneously generates both the wavenumber at which the fictitious thermostat begins and its temperature.  

\begin{figure}[ht!]
\begin{center}
 \hspace{-0.cm} \includegraphics[width=9.5cm,height=11.cm,angle=0]{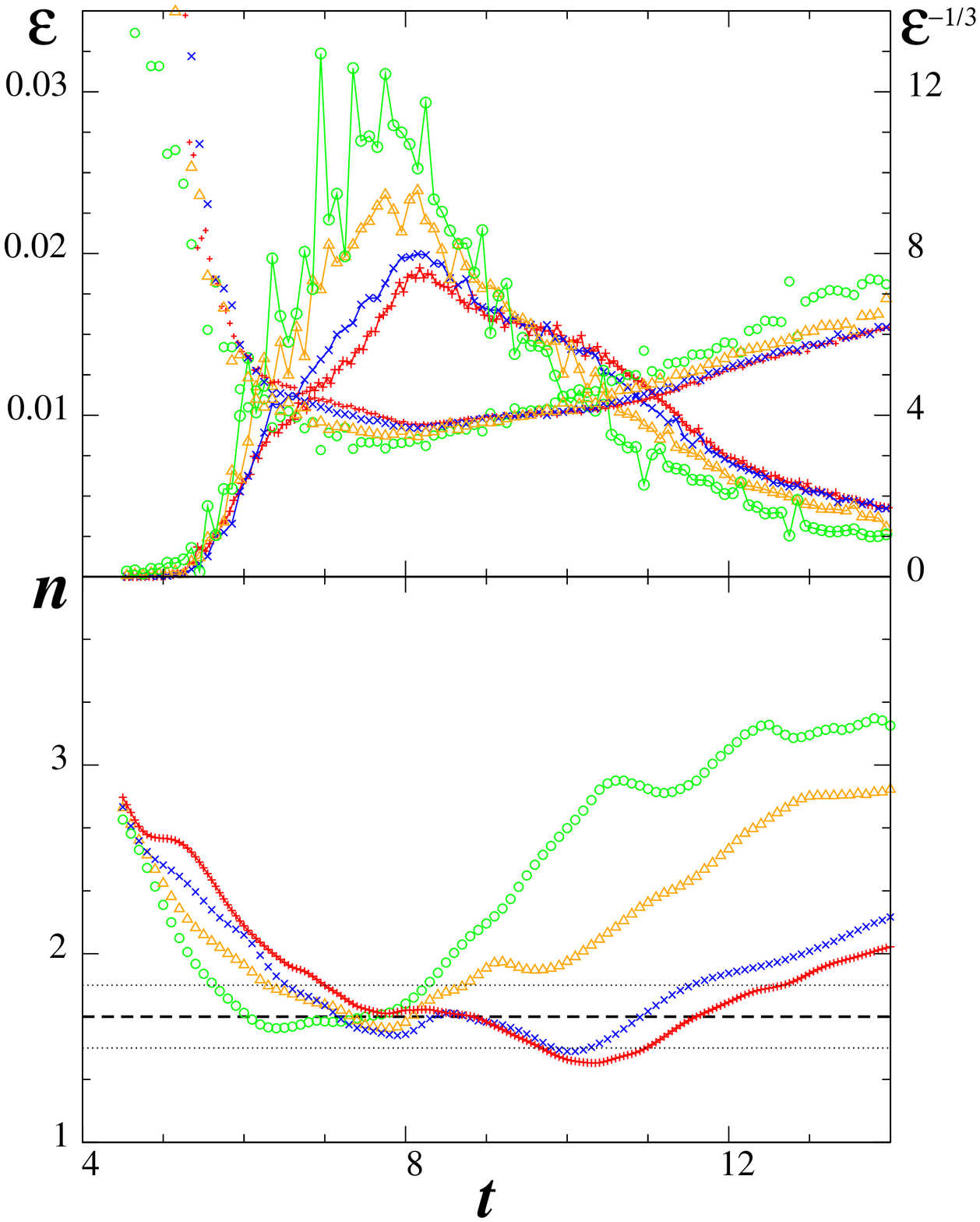}
\vspace{-1.5cm} 
\caption{\small{Temporal evolution of, top: energy dissipation $\varepsilon$ (left vertical axis) and $\varepsilon^{-1/3}$ (right vertical axis); bottom: $k^{-n}$ inertial range exponent $n$ at resolutions $256^3$ (fit interval $2\leq k \leq12$, circle $\circ$), $512^3$, (fit interval $2\leq k \leq14$, triangle $\triangle$), $1024^3$ (fit interval $2\leq k \leq16$, cross $\times$ and $1600^3$ (fit interval $2\leq k \leq20$, cross $+$).}
\label{tPd}}
\end{center}
\end{figure}

The previous results indicate scale separation between conservative large-scale and dissipative small-scale dynamics. Furthermore the scale separation increases with resolution. This
strongly suggests that large-scale behavior may be identical to that of high-Reynolds number standard Navier-Stokes equations, which is known \cite{coursuriel} to obey (at least approximately) K41 scaling.

The energy dissipation rate (\ref{Eps_def})
shown on figure \ref{tPd} (top, left axis) is in good agreement with the corresponding data
for the Navier-Stokes TG flow (see reference \cite{Brachet1},
figure 7 and reference \cite{coursuriel}, figure 5.12).
Both the time for maximum energy dissipation $t_{\rm max} \simeq 8$
and the value of the dissipation rate at that time
$\varepsilon(t_{\rm max}) \simeq 1.5 \,10^{-2}$ are in quantitative
agreement. 
Furthermore the long-time quasi-linear behavior of $\varepsilon^{-1/3}$ (shown on right axis) is compatible with K41 self-similar decay $\varepsilon(t)\sim L_0^{2} t^{-3}$.

A confirmation for K41 behavior around $t_{\rm max}$ is displayed on figure \ref{tPd} (bottom). The value of the inertial-range exponent $n$, obtained by low-$k$ least square fits of the logarithm of the energy spectrum to the function $c - n \log(k)$, is close to $5/3$ (horizontal dashed line) when $t \simeq t_{\rm max}$.
The $-5/3$ exponent is also shown as the left dashed line on bottom of figure \ref{SpecTh},
where the dissipative effects can be traced back to the energy spectrum decreasing faster than 
$k^{-5/3}$ at intermediate wavenumbers.

The mixed K41/absolute equilibrium spectra have already
been discussed in the wave turbulence literature ({\sl e.g.},\cite{Zakh}) and have more
recently been studied in connection with the Leith model of
hydrodynamic turbulence \cite{Colm}. In this context, small-scale thermalization may have some bearing on the so-called bottleneck problem if the dissipation wavenumber approaches $k_{\rm max}$.

Note that the dynamics of spectrally truncated  time-reversible nonlinear equations has also been 
investigated in the special cases of 1-D Burgers-Hopf models \cite{Majda1} and 
2-D quasi-geostrophic flows \cite{Majda2}.  
A central point in these studies was the nature of the statistical equilibrium that is achieved at 
large times. Several equilibria are {\it a priori} possible because both (truncated) 
1-D Burgers-Hopf and  2-D quasi-geostrophic flow models admit, besides the energy, a number of additional conserved quantities. 
The 3-D Euler case is of a different nature because (except for helicity that identically vanishes for the flows considered here) there is no known additional conserved quantity \cite{coursuriel} and the equilibrium is thus unique. The central problem in truncated 3-D Eulerian dynamics is therefore the mechanism of relaxation towards equilibrium, as studied in this letter. 

In summary, our main result is that the spectrally truncated Euler equation has long-lasting transients behaving just like those of the dissipative Navier-Stokes equation. The small-scale thermalized modes act as a fictitious microworld providing an effective viscosity to the large-scale modes. These dissipative effects were estimated using a new exact result based on Fluctuation Dissipation relations.
Furthermore, the solutions of the truncated Euler equations were shown to obey, at least approximately, K41 scaling.
In this context, the spectrally truncated Euler equations appears as a minimal model of turbulence.

{\bf Acknowledgments:}
We acknowledge discussions with D. Bonn, U. Frisch and Y. Pomeau. The computations were carried out on the NEC-SX5
computer of the Institut du D\'eveloppement et des Ressources en
Informatique Scientifique (IDRIS) of the Centre National pour la Recherche
Scientifique (CNRS).


\end{document}